\documentclass[12pt]{iopart}
\usepackage{hyperref}
\usepackage{graphicx}
\usepackage{caption}
\usepackage{booktabs}

\begin{document}

\title{Thermal formation of carbynes}

\author{Tobia M. Mazzolari$^1$ and Nicola Manini$^{1,2}$}
\address{
$^1$Dipartimento di Fisica, Universit\`a di Milano,
Via Celoria 16, 20133 Milano, Italy\\
$^2$CNR-IOM Democritos National Simulation Center, Via Bonomea
  265, 34136 Trieste, Italy
}
\ead{tobiamazzo@gmail.com, nicola.manini@mi.infm.it}

\begin{abstract}
We simulate the formation of $sp$ carbon chains (carbynes) by thermal
decomposition of $sp^2$ carbon heated by a hot discharge plasma, by means
of tight-binding molecular dynamics.
We obtain and analyze the total quantity of carbynes and their length
distribution as a function of temperature and density.
\end{abstract}



\section{Introduction}

The $sp$ form of carbon, carbyne, has proved more elusive than its $sp^3$
(diamond) and $sp^2$ (graphite/graphene) counterparts.
Recently, also in view of the potential exciting applications
\cite{Akdim11, Cretu13, Guo13, Zanolli10, Zanolli11, Zeng10, Erdogan11,
  Liu13} $sp$ carbon chains (spCCs, or carbynes) are being produced in
significant amount and investigated extensively.
Besides synthetic routes \cite{Mohr03, Zhao03, Cataldo05, Chalifoux09,
  Rice10, Cataldo10} leading to spCCs bonded to molecular ligands
and various mechanical or electromagnetic methods \cite{Jin09, Chuvilin09,
  Mikhailovskij09, Mazilova10, Inoue10} addressing the single spCC, carbyne
has been formed in substantial amounts (together with other carbon
clusters) via homogenous reactions within the hot plasma produced by an
electric discharge at the surface of a graphite electrode
\cite{Barborini99a,Barborini99b, Piseri01, Ravagnan02,Casari04,Ravagnan07}.
Despite their high reactivity, a significant fraction of these carbynes
survive landing on a solid surface, where they are typically stabilized by
grafting to other carbon clusters, mainly of $sp^2$ hybridization
\cite{Castelli12}.
After deposition, carbynes are detected in the nanostructured film by
various kinds of spectroscopies
\cite{Ravagnan07,Ravagnan09,Cinquanta11,Ravagnan11}.

In the present work we investigate the early formation stages of spCCs at
the interface of a graphite electrode with a hot plasma.
Specifically, we investigate the length distribution $P(m)$ of C$_m$
chains, and its dependence on the temperature of the hot plasma and on the
local density of the available carbon material.
The time evolution of $P(m)$ and its temperature dependence can provide
information useful for the tuning of the plasma properties, in view of an
optimization of the formation of carbynes of a desired length.
A previous simulation work \cite{Yamaguchi07} used a similar model to
investigate the transformation of $sp$ structures into ordered structure,
such as fullerenes and nanotubes under thermal annealing.
Here we focus instead on a quantitative characterization of the initial
formation of the carbynes, in a high-density, high-temperature plasma
plume.

\section{Methods}

A reliable description of the $sp^2 \to sp$ interconversion requires a
transferable description of carbon binding in its different hybridizations.
In view of these requirements, we adopt a well-established tight-binding (TB)
model for the adiabatic potential of a carbon-only material \cite{Xu92}.
In \ref{validation:sec} we validate the adopted model against {\it
  ab-initio} density functional theory (DFT) simulations, in the local
spin-density approximation (LSDA).
As is standard in the literature of the field
\cite{Yamaguchi07,Bonelli09,Cadelano09,Ortolani12}, we evaluate the
Hellmann-Feynman forces \cite{Colombo05} for an efficient simulation of the
classical time evolution of the ions of the carbon material.

\begin{figure}
\centering
\includegraphics[width=.55\textwidth,clip=]{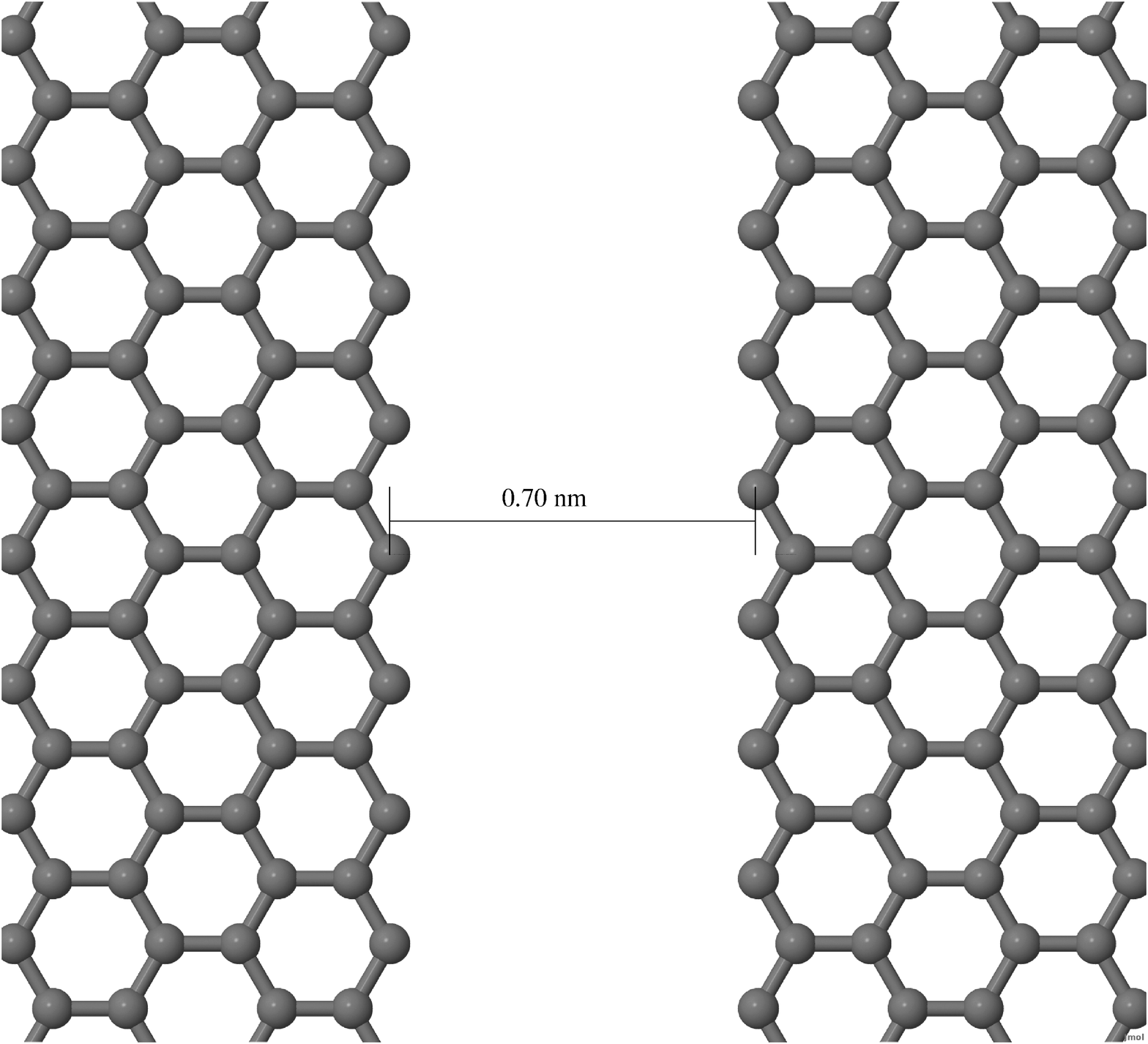}
\caption{\label{grafene-ribbon:fig}
Starting configuration of the $T=0$ carbon ribbon, obtained by selecting a
portion of an ideal graphene sheet.
Periodic boundary conditions are applied over a parallelepiped box in all
three orthogonal directions.
The $\sim 0.70$~nm-wide central cut exposes the ribbon zig-zag edges.
}
\end{figure}

We adopt a graphene nanoribbon to represent the initial state of the
$sp^2$-carbon sacrificial electrode.
A graphitic multilayer would make little difference because the adopted TB
force field is rather short ranged and does not include Van-der-Waals
interactions.
The infinite nanoribbon is represented by a $sp^2$ carbon stripe inside a
parallelepiped simulation cell, to which we apply periodic boundary
conditions (PBC), see Fig.~\ref{grafene-ribbon:fig}.
We simulate the heat exchanges of the sample with the hot plasma and the
bulk of the carbon electrode by means of a standard Langevin thermostat
\cite{Allen91}.
The equation of motion of each carbon atoms is
\begin{equation}
\label{eqn:DLangevin}
\frac {d\mathbf{p}_i}{dt} =
-\eta \, \mathbf{p}_i(t) + \mathbf{f}_i(t) + \mathbf{\tilde{f}}_i(t)
\,,
\end{equation}
where $\mathbf{p}_i$ is the momentum of atom $i$, $\mathbf{f}_i(t)$ is the
(Hellmann-Feynman) force acting on atom $i$ due to the interaction with all
other atoms in the sample, $\mathbf{\tilde{f}}_i(t)$ is a
Gaussian-distributed stochastic force representing the collisions of atom
$i$ with the hot plasma, and $\eta$ is the phenomenological parameter
accounting for the rate of heat exchange with the thermostat
\cite{Allen91}.
We integrate these equations numerically with a $0.5$~fs time step.
The other simulation details are summarized in Table~\ref{tab:sim}.

\begin{table}
\caption{\label{tab:sim}
The general conditions used for the considered groups of simulations.
The density is changed by varying the size of the simulation box in the
direction perpendicular to the plane initially containing the nanonribbon.
}
\centering
\begin{tabular}{cccccccc}
\toprule
$\eta$		& $T$	& edge & number  &  density	 & total time	& number of\\
 $[$ps$^{-1}]$     &[K]	& type & of atoms&[atoms nm$^{-3}$]& [ps] 	& simulations\\
\midrule
0.1 & 4000 & zig-zag & 112 & 2.77  & 150 & 5\\
0.1 & 4500 & zig-zag & 112 & 2.77  & 150 & 5\\
0.1 & 5000 & zig-zag & 112 & 2.77  & 250 & 5\\
0.1 & 6000 & zig-zag & 112 & 2.77  & 150 & 5\\
1   & 4000 & zig-zag & 112 & 2.77  & 50  & 10\\
1   & 4000 & zig-zag & 112 & 9.23  & 500 & 10\\
1   & 4500 & zig-zag & 112 & 2.77  & 50  & 10\\
1   & 4500 & zig-zag & 112 & 9.23  & 200 & 10\\
1   & 5000 & zig-zag & 112 & 1.39  & 100 & 10\\
1   & 5000 & zig-zag & 112 & 2.77  & 100 & 10\\
1   & 5000 & zig-zag & 112 & 9.23  & 100 & 10\\
1   & 5000 & zig-zag & 112 & 13.85 & 100 & 10\\
1   & 5000 & zig-zag &  56 & 9.23  & 100 & 10\\
1   & 5000 & armchair&  96 & 9.23  & 100 & 10\\
1   & 6000 & zig-zag & 112 & 2.77  & 50  & 10\\
1   & 6000 & zig-zag & 112 & 9.23  & 100 & 10\\
5   & 4000 & zig-zag & 112 & 2.77  & 50 & 5\\
5   & 4500 & zig-zag & 112 & 2.77  & 50 & 5\\
5   & 5000 & zig-zag & 112 & 2.77  & 50 & 5\\
5   & 6000 & zig-zag & 112 & 2.77  & 50 & 5\\
10  & 4000 & zig-zag & 112 & 2.77  & 50 & 5\\
10  & 4500 & zig-zag & 112 & 2.77  & 50 & 5\\
10  & 5000 & zig-zag & 112 & 2.77  & 50 & 5\\
10  & 6000 & zig-zag & 112 & 2.77  & 50 & 5\\
\bottomrule
\end{tabular}
\end{table}

In the experiment, an electric discharge ionizes a short injected He-gas
pulse.
Ablation occurs when helium plasma strikes the carbon cathode surface,
removing atoms via sputtering \cite{Barborini99a,Barborini99b,
  Piseri01,Ravagnan02}.
The duration of the plasma pulse, in the sub-millisecond region
\cite{Piseri01}, is orders of magnitude longer than typical sub-nanosecond
times we can afford to explore in simulation.
Over the nanosecond time scale, an essentially steady sputtering regime has
established, with the progressive erosion of the bulk graphite electrode,
and with the eroded carbon material diffusing away from the electrode
through the hot plasma.
Local thermal equilibrium is likely to be maintained by frequent collisions
with the plasma, and successive progressive cooling is achieved over a time
scale of microseconds, mainly by fragmentation and emission of
electromagnetic radiation.
In our atomistic model anything comparable to this steady decomposition
regime is out of reach due to the huge number of atoms and far too long
simulations times required.
We rather gather information relevant to the experimental regime by
investigating the transient decomposition of the nanoribbon when suddenly
brought to high temperature.

Within the fixed simulation volume and at high temperature, the simulated
sample progressively decomposes into isolated atoms, dimers, longer
carbynes C$_m$, and in principle also $sp^2$ fragments, although we observe
few of the latter within the simulated conditions.
In simulation, erosion proceeds until the $sp^2$ nanoribbon is completely
decomposed and a final steady regime is reached, which can be described as
a hot atomic/molecular carbon gas.

\begin{figure}
\centerline{
\hfill\includegraphics[width=.4\textwidth,clip=]{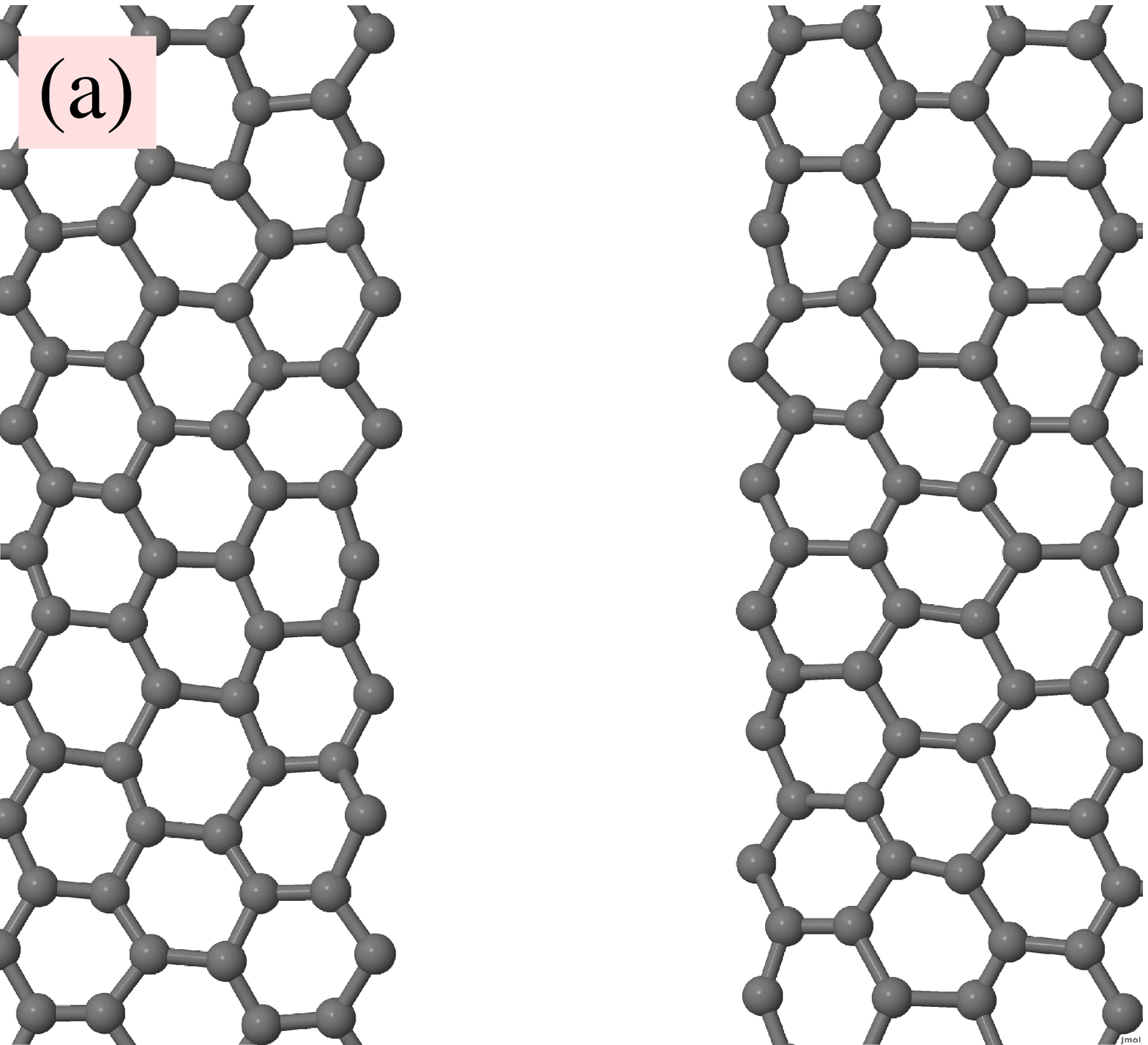} \hfill
\includegraphics[width=.4\textwidth,clip=]{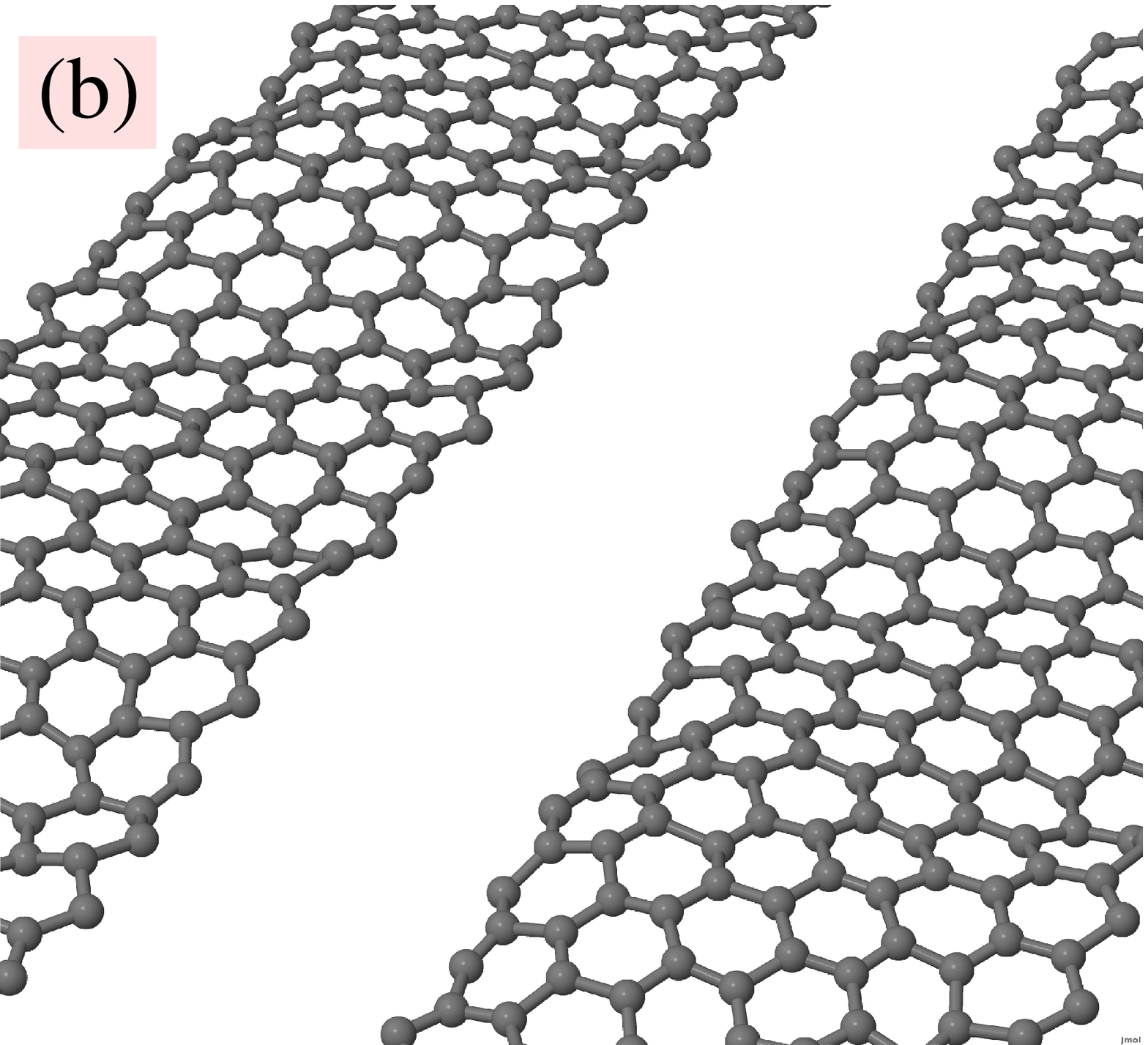}\hfill
}
\caption{\label{grafene-initial:fig}
Initial configuration prepared starting from the ideal nanoribbon
configuration of Fig.~\ref{grafene-ribbon:fig} evolved for $4$~ps at an
intermediate temperature $T=2000$~K.
(a) A top view of a single supercell; (b) an inclined perspective view of
the infinitely repeated-cell geometry.
When this snapshot is adopted as the starting point of successive
simulations, the initial velocities are randomly reset according to a
thermal distribution appropriate to the new simulation temperature.
The PBC parallelepiped box remains unchanged, thus fixing the overall
atomic density.
}
\end{figure}

The dynamics of spCC formation can be addressed precisely by analyzing the
transient regime where the nanoribbon gets eroded.
To study this transient we adopt the following protocol:
(i) we start with an ideal $T=0$ configuration such as the one depicted
in Fig.~\ref{grafene-ribbon:fig};
(ii) we run a pre-thermalization simulation of $4$~ps at the intermediate
temperature $T=2000$~K, obtaining the configuration represented in
Fig.~\ref{grafene-initial:fig} where a significant thermal excitation is
already present in the phononic degrees of freedom, but no bonds are
broken;
(iii) we then attribute random initial velocities to the atoms, from a
Maxwellian distribution at the desired simulation temperature, see
Table~\ref{tab:sim}, and start off the ``production'' Langevin run.

To generate a fair statistics, for given physical conditions we repeat the
Langevin evolution at least 5, but usually 10 times, starting from the same
initial configuration but with independent random sets of initial
velocities and stochastic forces $\mathbf{\tilde{f}}_i(t)$.

\section{Results}

\begin{figure}
\centering
\includegraphics[width=.8\textwidth,clip=]{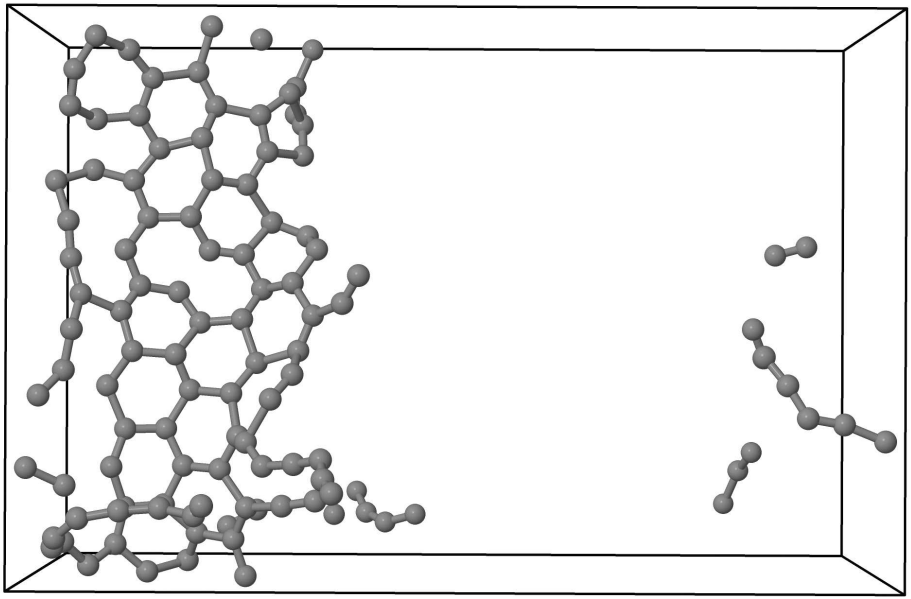} \hfill
\caption{\label{snapshot:fig}
A typical snapshot of the  $T=5000$~K evolution of the sample starting off
in the configuration of Fig.~\ref{grafene-initial:fig}.
Note an isolated C atom (top left), a dimer (mid-height right), a C$_3$
(bottom right), a C$_4$ (bottom center), a C$_8$ chain crossing the box PBC
and a few carbynes still partly attached to the $sp^2$ region.
}
\end{figure}

The starting configuration is equilibrated, with both kinetic (velocities)
and potential (positions) degrees of freedom representing the appropriate
Boltzmann energy distribution at $T=2000$~K.
As soon as the temperature is turned up to the ``production'' value listed
in Table~\ref{tab:sim}, the thermostat immediately transfers energy to the
kinetic degrees of freedom.
Subsequently, this extra energy is shared very rapidly with the potential
degrees of freedom of the sample.
This increased potential energy results in the progressive decomposition of
the $sp^2$ sample, with the detachment of isolated atoms, clusters, and
$sp$ chains of different lengths.
We simulate temperatures which are rather high compared to those which
could be estimated in experiment, because they have the advantage to speed
up the decomposition dynamics substantially, and bring it down to an
accessible $10$~ps time scale.
At the considered temperatures and densities, isolated atoms and $sp$ chains
are the dominant emitted clusters.
As an example, Fig.~\ref{snapshot:fig} illustrates a typical snapshot where
a C atom, C$_2$, C$_3$, C$_4$ chains, and a C$_8$ chain crossing the PBC
are visible.
All these carbynes are fully detached from the $sp^2$ nanoribbon.
The main spCC formation mechanism involves random self-cleavage of the
graphene edge, as visible e.g.\ at the left edge of the graphene sheet in
Fig.~\ref{snapshot:fig}.

\subsection{The carbyne length distribution}

\begin{figure}
\centerline{
 \includegraphics[width=.55\textwidth,clip=]{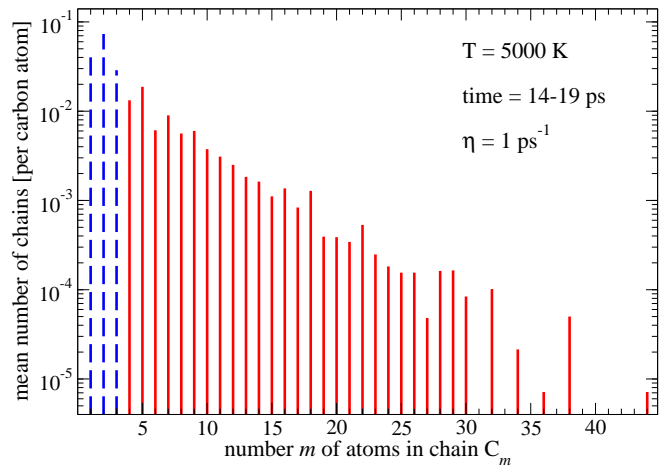}
}
\caption{\label{mean_number_of_chains:fig}
(Color online)
Histogram of the frequency of individual spCC lengths.
Data are averaged over 500 equally-spaced snapshots in the time interval
from $14$ to $19$~ps of the evolution started when turning on a $T=5000$~K
Langevin thermostat characterized by a coupling rate $\eta=1$~ps$^{-1}$.
Further averaging is realized over 10 independent Langevin trajectories.
The mean atomic density is $n=9.23$~nm$^{-3}$.
}
\end{figure}

Figure~\ref{mean_number_of_chains:fig} reports the distribution of length
(expressed as number of atoms $m$) of the detaching C$_m$ carbynes.
This histogram includes isolated atoms and free chains and also carbynes
under formation, still bound to the graphene edge at either or both ends.
The $m=1$ column counts both isolated atoms and atoms attached with one
bond to the graphene sheet.
$sp^2$ ($sp^3$) atoms forming 3 (4) bonds do not contribute to this
statistics.
To evaluate this histogram, we have developed a computer code which, taking
PBC into account, identifies $sp$ atoms by being connected by either one or
two bonds, and then follows the bonds and characterizes uniquely each
carbyne.

The obtained distribution is dominated by single atoms, dimers and very
short chains (dashed columns).
For longer chains, i.e.\ proper carbynes, the length distribution decays
approximately exponentially as $m$ increases.
A small but detectable fraction extends to quite long carbynes with $m\geq
20$.
A weak even-odd unbalance is visible.

\begin{figure}
\centerline{
 \includegraphics[width=.55\textwidth,clip=]{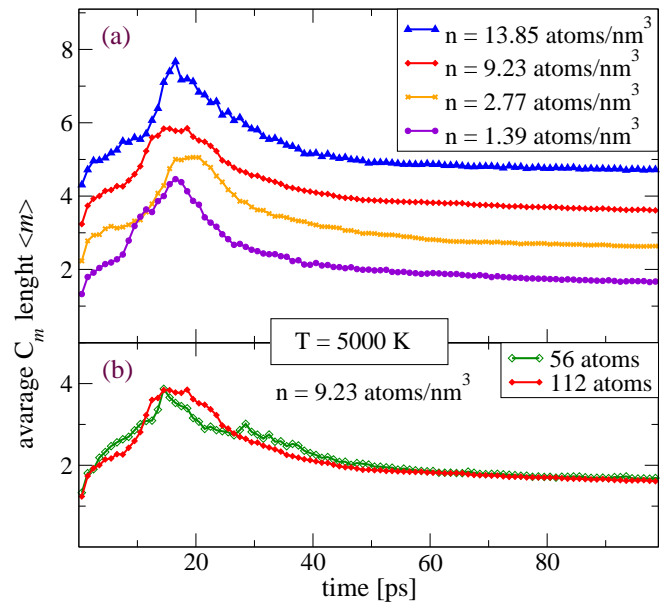}
}
\caption{\label{average_c_m_length_m:fig}
(Color online)
The time evolution of the average length $\langle m\rangle$ of spCC from
the simulation start at $T=5000$~K, averaged over 10 independent
simulations and over 100 samples within the same ps.
(a) Successive curves at different simulation-box volume, i.e.\ different
mean atomic density, are mutually shifted upward by 1 for clarity.
(b) Size effects illustrated by the comparison of the data for density
$9.23$~atoms$/$nm$^3$ with similar simulations (open symbols) carried out
with a sample composed of one half of the atoms in one half volume.
}
\end{figure}

\begin{figure}
\centerline{
 \includegraphics[width=.55\textwidth,clip=]{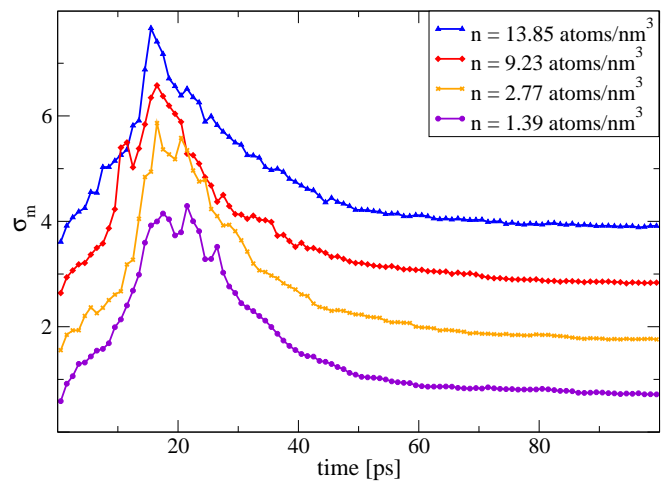}
}
\caption{\label{sigma:fig}
(Color online)
The time evolution of the standard deviation of the spCC length
distribution, as in Fig.~\ref{average_c_m_length_m:fig}.
Successive curves at increasing mean atomic densities are shifted upward by
$1$ for readability.
}
\end{figure}

The distribution depicted in Fig.~\ref{mean_number_of_chains:fig}
represents a snapshot of an evolving distribution, and precisely as it
occurs approximately 15~ps after turning on the interaction with the
high-temperature environment.
To characterize the time evolution of the chain-length distribution,
Figs.~\ref{average_c_m_length_m:fig}(a) and \ref{sigma:fig} report
respectively the average spCC length $\langle m\rangle$ and its fluctuation
$\sigma_m = (\langle m^2\rangle-\langle m\rangle^2)^{1/2}$, as a function
of time.
The mean carbyne length $\langle m\rangle$ is seen to peak after
approximately 20~ps of interaction with the hot thermostat, at a peak value
$\langle m\rangle\simeq 4$.
Successively, long spCCs decay rather rapidly, and the sample decays to
mainly isolated atoms and dimers, with very few longer carbynes.
The width $\sigma_m$ of the distribution, Fig.~\ref{sigma:fig}, follows a
similar pattern, and is also especially broad (of the same order as
$\langle m\rangle$) in the 20~ps region.

We make sure that size effects are negligible by repeating similar
calculations for a sample of one half the standard size, obtained by
cutting the nanoribbon in Fig.~\ref{grafene-initial:fig} horizontally.
The evolution of $\langle m\rangle$ shown in
Fig.~\ref{average_c_m_length_m:fig}(b) shows little or no size effects.
The detailed chain length distribution of the smaller-size simulation (not
shown) is very similar to that of Fig.~\ref{mean_number_of_chains:fig}, but
for a significant depletion of the long-chain region.

\begin{figure}
\centerline{
 \includegraphics[width=.55\textwidth,clip=]{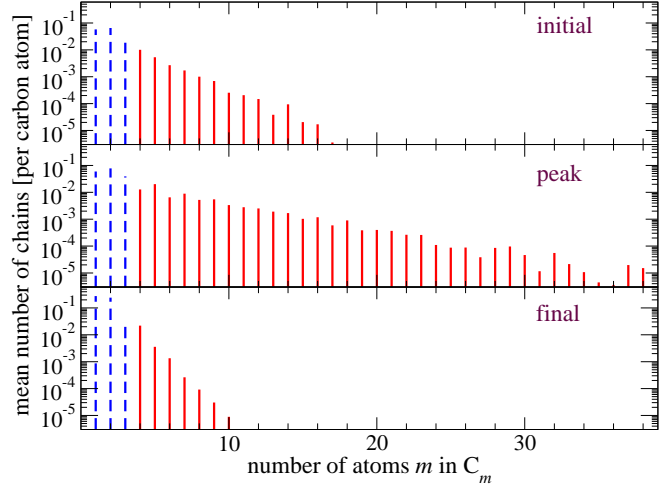}
}
\caption{\label{mean_number_of_chains_three_times:fig}
(Color online)
Successive distributions of C$_m$ lengths.
The time intervals are: initial --- $0\div 10$~ps; peak --- $15\div 25$~ps;
final --- $90\div 100$~ps.
The simulation and statistical parameters are the same as in
Fig.~\ref{mean_number_of_chains:fig}.
}
\end{figure}

This evolution is illustrated also by the three histograms representing the
initial, peak, and final time intervals,
Fig.~\ref{mean_number_of_chains_three_times:fig}.
We see that monomers and dimers dominate the distribution at all times, and
especially  in the long-time limit.
Long carbynes are present as a significant fraction mainly in the $\langle
m\rangle$-peak region, i.e.\ around 20~ps after starting to heat at
$T=5000$~K.
As we are interested in the formation of proper carbynes, from now on we
focus on spCCs C$_m$ with $m\geq 4$, namely those marked by solid lines in
Figs.~\ref{mean_number_of_chains:fig} and
\ref{mean_number_of_chains_three_times:fig}, with the understanding that
these are a minority fraction of the entire emitted carbon material.

\begin{figure}
\centerline{
 \includegraphics[width=.55\textwidth,clip=]{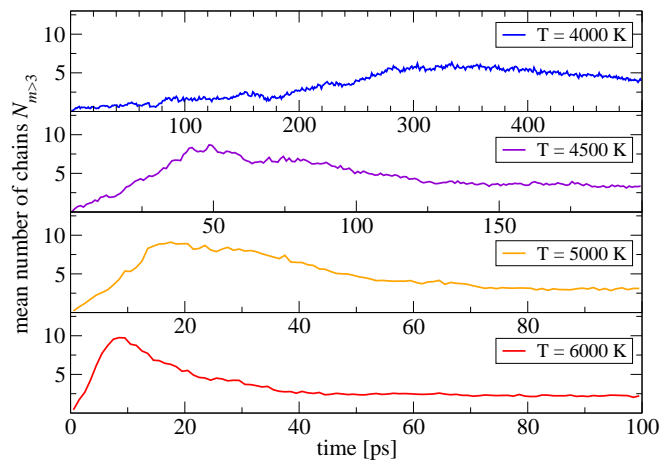}
}
\caption{\label{chains_dependence_from_temperature:fig}
(Color online)
The time evolution of the number of carbynes C$_m$ with $m\geq 4$ atoms.
Each curve at a different thermostat temperature is the result of averaging
over 10 independent simulations and over 100 snapshots within each
picosecond.
The considered atomic density here is $n=9.23$~nm$^{-3}$.
}
\end{figure}

To investigate how the spCCs production depends on temperature, we run and
compare multiple simulations at different temperatures.
Figure~\ref{chains_dependence_from_temperature:fig} illustrates the
evolution of the mean number $N_{m>3}$ of spCCs involving 4 atoms or more,
for four different temperatures.
The general trend of $N_{m>3}$ exhibits an initial increase, a broad peak,
eventually followed by a systematic decay, regardless of temperature.
Note however that the horizontal time scale is far more extended for the
simulations carried out at lower temperature, where all decomposition
phenomena occur over a far longer time scale.
In the final region
most $sp^2$ material has undergone a radical deterioration, and the sample
has reached the state of an essentially gaseous mixture.
The trend reported in Fig.~\ref{chains_dependence_from_temperature:fig}
suggest that an optimal production rate of carbyne can be achieved provided
that carbynes move out of the hot region in a time long enough for a
significant concentration of carbynes to arise, but also short enough that
the successive decomposition of these carbynes to dimers and isolated atoms
has not proceeded significantly.
This optimal time becomes shorter and shorter as the plasma temperature is
raised.

\subsection{The intrinsic formation rate}

The formation rate of spCCs does not depend only on temperature but also on
the effectiveness of the contact between the carbon sample and the hot
plasma.
In our model, we represent this contact by the strength $\eta$ of the
coupling between the system and the thermostat.
In experiment, this coupling is tuned by such properties as the local
pressure, flux, and atomic mass of the hot plasma.

In addition to this ``extrinsic'' rate, we expect a ``natural''
contribution to the rate of formation of spCCs related to the intrinsic
kinetic mechanisms involved in the erosion of the carbon material, and
independent of the details of the coupling to the plasma/thermostat.
To extract this ``natural'' formation rate of carbynes to $dN_c/dt \equiv
dN_{m>3}/dt$ , we need to subtract the thermostat contribution.
To this purpose, we simulate and compare the same conditions, with varied
thermostat transfer rate $\eta$.

\begin{figure}
\centerline{
\includegraphics[width=.55\textwidth,clip=]{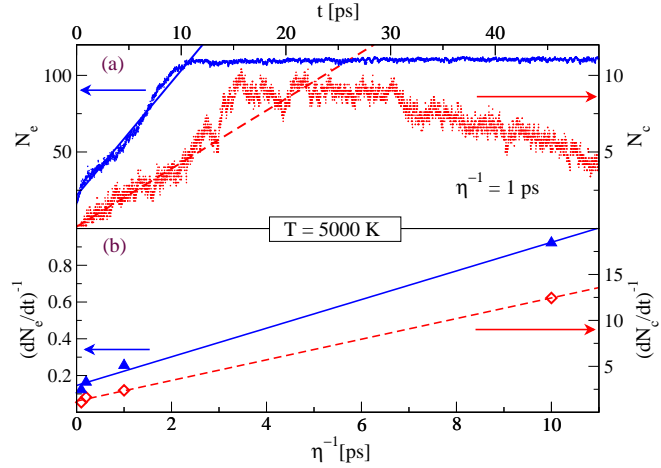}
}
\caption{\label{fit:fig}
(Color online)
The determination of the intrinsic generation rate for carbynes ($\tau_c$)
and for eroded carbon atoms ($\tau_e$).
(a) Firstly, the decomposition rate $dN_e/dt$ is determined as the slope of
the linear fit of the initial raising time interval of a simulation.
(b) Following Eq.~(\ref{tau:eq}). the decomposition rates obtained for
different thermostat coupling $\eta$ are plotted as a function of
$\eta^{-1}$, and a further linear fit is carried out.
The intercept resulting from these second fits provide precisely the best
estimate of the intrinsic $\tau_e$ (and similarly $\tau_c$) for the
considered simulation temperature $T$, as reported in Table~\ref{tab:tau}.
}
\end{figure}

\begin{table}
\caption{\label{tab:tau}
The intrinsic formation time for the generation of carbynes ($\tau_c$) and
for detached carbon atoms ($\tau_e$) in any form (including carbynes).
These data are obtained by a double fitting procedure illustrated in
Fig.~\ref{fit:fig}.
} \centering
\begin{tabular}{ccc}
\toprule
Temperature [K] & $\tau_c$ [ps] & $\tau_e$ [ps]\\
\midrule
4500 & 4.22$\pm$0.31  & 0.44$\pm$0.02\\
5000 & 1.22$\pm$0.14  & 0.15$\pm$0.02\\
6000 & 0.19$\pm$0.04 & 0.025$\pm$0.005\\
\bottomrule
\end{tabular}
\end{table}

The detachment of one or multiple carbynes is an endothermic process
requiring the concentration of energy at some point of the sample.
This barrier energy $E_B$, in the eV range, is provided at the expense of
the sample thermal energy, and it is eventually restored to the whole
sample by the thermostat.
This scheme configures the ``series'' of two processes occurring at
definite rates: (i) the carbyne formation itself, at an intrinsic rate
$\tau_c^{-1}$, and (ii) the transfer of heat from the thermostat, at a rate
$\eta$.
As these processes are not concurrent (in parallel) but sequential (in
series, like chains of radioactive decays), the overall rate $dN_c/dt$
satisfies
\begin{equation}\label{tau:eq}
\left(\frac{dN_c}{dt}\right)^{-1} = D\,\eta^{-1} + \tau_c
\,,
\end{equation}
where $D$ is a dimensionless coefficient of order unity.
$dN_c/dt$ is extracted by fitting the initial linearly-increasing region of
the computed $N_c$ as a function of time, as in
Fig.~\ref{chains_dependence_from_temperature:fig}, and specifically
Fig.~\ref{fit:fig}a.
Given these values of $dN_c/dt$ computed at varied $T$ and $\eta$, we
extract the intrinsic formation time $\tau$ for a spCC by fitting the
observed $(dN_c/dt)^{-1}$ as a function of $\eta^{-1}$.
The data are compatible with a straight line, see Fig.~\ref{fit:fig}b.
According to Eq.~(\ref{tau:eq}), the intercept at $\eta^{-1}=0$ provides
precisely $\tau_c$.
We repeat this procedure for different temperature, and obtain the
intrinsic formation time $\tau_c$, listed in Table~\ref{tab:tau}.

\begin{figure}
\centerline{
\includegraphics[width=.55\textwidth,clip=]{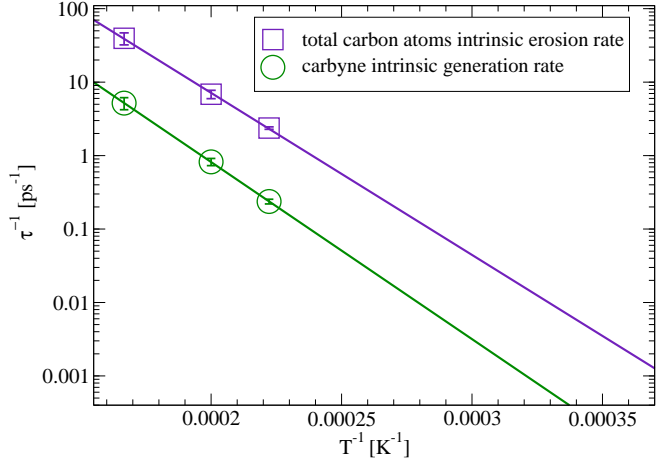}
}
\caption{\label{arrhenius:fig}
(Color online)
The activated temperature dependence of the intrinsic generation rate of
the chains $\tau^{-1}$.
The data points are those reported in Table~\ref{tab:tau}, the solid line
is a best fit.
}
\end{figure}

Figure~\ref{arrhenius:fig} reports the obtained rates $\tau_c^{-1}$ as a
function of the inverse temperature.
The variation of the intrinsic formation rate $\tau_c^{-1}$ is compatible
with an Arrhenius law
\begin{equation}
\log(\tau_c^{-1}) = \log(\tau_{c\,\infty}^{-1}) - \frac{E_B}{k_{\rm B}T}
\,.
\end{equation}
The estimated barrier $E_B= 4.78\pm 0.02$~eV, of course in agreement with
the carbyne breakup energy of Fig.~\ref{compare:fig} in \ref{validation:sec}.
The attempt rate $\tau_\infty^{-1}\simeq 5.4\times 10^{16}$~s$^{-1}$ is
quite large compared to typical attempt rates of molecular chemical
reactions.
The reason is that this rate refers to the entire sample.

An analogous analysis carried out for the total number $N_e$ of eroded
carbon atoms (also illustrated in Figs.~\ref{fit:fig} and
\ref{arrhenius:fig}) yields a similar barrier (4.38~eV) and an
approximately double attempt rate $\tau_{e\,\infty}^{-1}\simeq 1.8\times
10^{17}$~s$^{-1}$.
We can assume that $\tau_c^{-1}$ and $\tau_e^{-1}$ are rates appropriate
for the length of $sp^2$ carbon edges exposed to the plasma in our
simulated sample (Fig.~\ref{grafene-initial:fig}), i.e.\ approximately
$4$~nm.
Since the inter-layer distance in graphite is $0.335~$nm, and accounting
from orientational disorder, our sample is equivalent to an exposed
erodible surface area of approximately $S\simeq 1$~nm$^2$.
At a plausible temperature next to the graphite surface, say 2500~K, the
extrapolated intrinsic rate $\tau_e^{-1}\simeq 3\times 10^8$~s$^{-1}$ is
compatible with an erosion rate
$(S\tau_e)^{-1} \simeq 3\times 10^{26}$~s$^{-1}$\,m$^{-2}$, corresponding
to $\sim 5$~kg\;s$^{-1}$\,m$^{-2}$.
This mass rate accounts for an erosion speed in the order of $\sim 2$~mm/s.
This figure is quite close to the electrode erosion speed observed during
the active plasma-pulse part of the source cycle in experiment
\cite{Barborini99a,Piseri01}.
%
Of this eroded carbon, the fraction of ``native'' $m > 3$ C$_m$ carbynes is
estimated to approximately $22$\% at the same temperature of $2500$~K
(assuming an average carbyne length of 5).

\subsection{Dependence on density and edge type}

\begin{figure}
\centerline{
 \includegraphics[width=.55\textwidth,clip=]{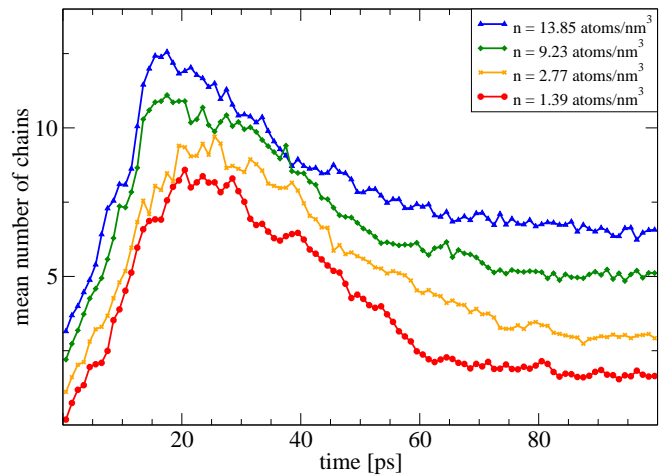}
}
\caption{\label{chains_dependence_from_density:fig}
(Color online)
The time evolution of the number of spCC longer than $m=3$ atoms.
Each curve at a different atomic density is the result of averaging over 10
independent simulations at $T=5000$~K and is shifted upward by 1 for
readability.
}
\end{figure}

It is necessary to examine how the amount and fraction of formed spCCs
depends on the overall carbon density in the formation region.
In the experimental conditions, the carbon density drops rapidly from that
of bulk graphite inside the rod to a dilute-gas figure downstream in the
carrier gas flux away from the surface.
As it is currently impractical to determine the precise atomic density in
the erosion region \cite{Barborini99a, Piseri01, Ravagnan07}, we explore a
range of densities.
The data of Fig.~\ref{chains_dependence_from_density:fig} (see also
Figs.~\ref{average_c_m_length_m:fig} and \ref{sigma:fig}) indicate that the
carbon density does not affect dramatically the carbyne formation.
Mostly, a higher density tends to increase slightly the early-time
formation rate of carbynes, thus anticipating the time at which the number
of formed carbynes is maximum.

\begin{figure}
\centerline{
 \includegraphics[width=.55\textwidth,clip=]{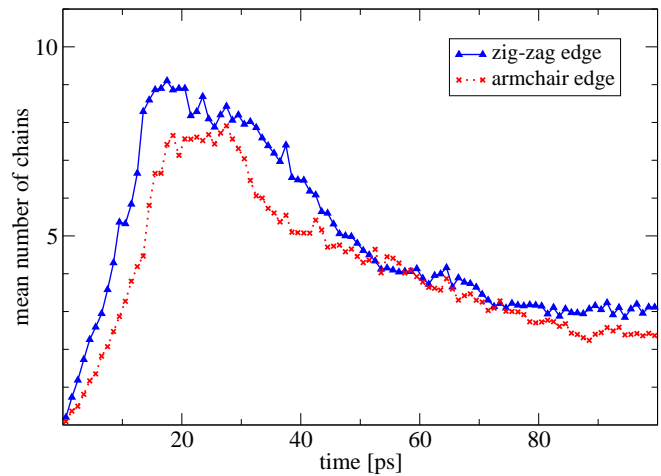}
}
\caption{\label{armchair_zigzag:fig}
(Color online)
Comparison of the number of spCC longer than $m=3$ atoms evolved from a
$sp^2$ sheet terminated with armchair vs. zig-zag edges.
Each curve is the result of averaging over 10 independent simulations at
temperature $T=5000$~K and the same atomic density $n=9.23$~nm$^{-3}$.
}
\end{figure}

Until now, all simulations represent solid carbon with the nanoribbon
characterized by a zig-zag edge, as in Figs.~\ref{grafene-ribbon:fig} and
\ref{grafene-initial:fig}.
To make sure that the carbyne formation rate is not affected by this
assumption, we simulate also a nanoribbon exposing an armchair edge.
Figure~\ref{armchair_zigzag:fig} reports the comparison of the carbyne
formation.
While the armchair edge proves significantly more stable, with a slower
initial rate of spCC formation, the observed difference is not especially
large, approximately a factor two.
We conclude that the explored zig-zag model is well suitable for estimating
the rates, which it probably slightly overestimates compared to a random
mix of zig-zag and armchair edges as in actual graphite.

\section{Discussion and conclusions}
%

The adopted  Langevin-thermostat model represents a substantial
idealization of the erosion of $sp^2$ carbon interacting with a hot
plasma.
The significant inhomogeneity and temperature gradient occurring in the
experimental setup are surely lacking in this model.
All the same, the exploration of a range of temperature and densities shows
that (i) temperature differences induce radical (activated) differences
mainly in the overall production rate, but not on the long-time statistical
properties of the formed carbynes; (ii) density variations induce minor
changes on the carbyne formation.
These results suggest that, despite its limitations, the adopted TB model
allows us to explore the carbon decomposition and carbyne formation in a
qualitative, but also statistically quantitative way.

In experiment, the carbynes and other produced clusters migrate into a
cooler region where they may in part undergo fragmentation, recombination,
and other transformations, including the synthesis of graphitic clusters,
fullerenes, nanotubes, and amorphous structures.
Contrary to the described carbyne formation, the recombination phenomena,
already studied in the past \cite{Yamaguchi07}, depend crucially on the
probability of encounter and quenching of primary clusters, which in turn
depends on the density of available carbon material.
Accordingly, also the carbyne depletion between their generation in the
hottest part of the plasma plume and the final deposition can occur at a
rate strongly dependent on the plasma-gas pressure and flux.
Further quantitative consideration of these phenomena may prove necessary
in view of the direct interest in the formation of carbynes.

\ack

We thank Nicola Ferri, Giovanni Onida, and Paolo Piseri for useful
discussion.

\appendix
\section{Validation of the TB potential}
\label{validation:sec}

\begin{figure}
\centering
\includegraphics[width=.55\textwidth,clip=]{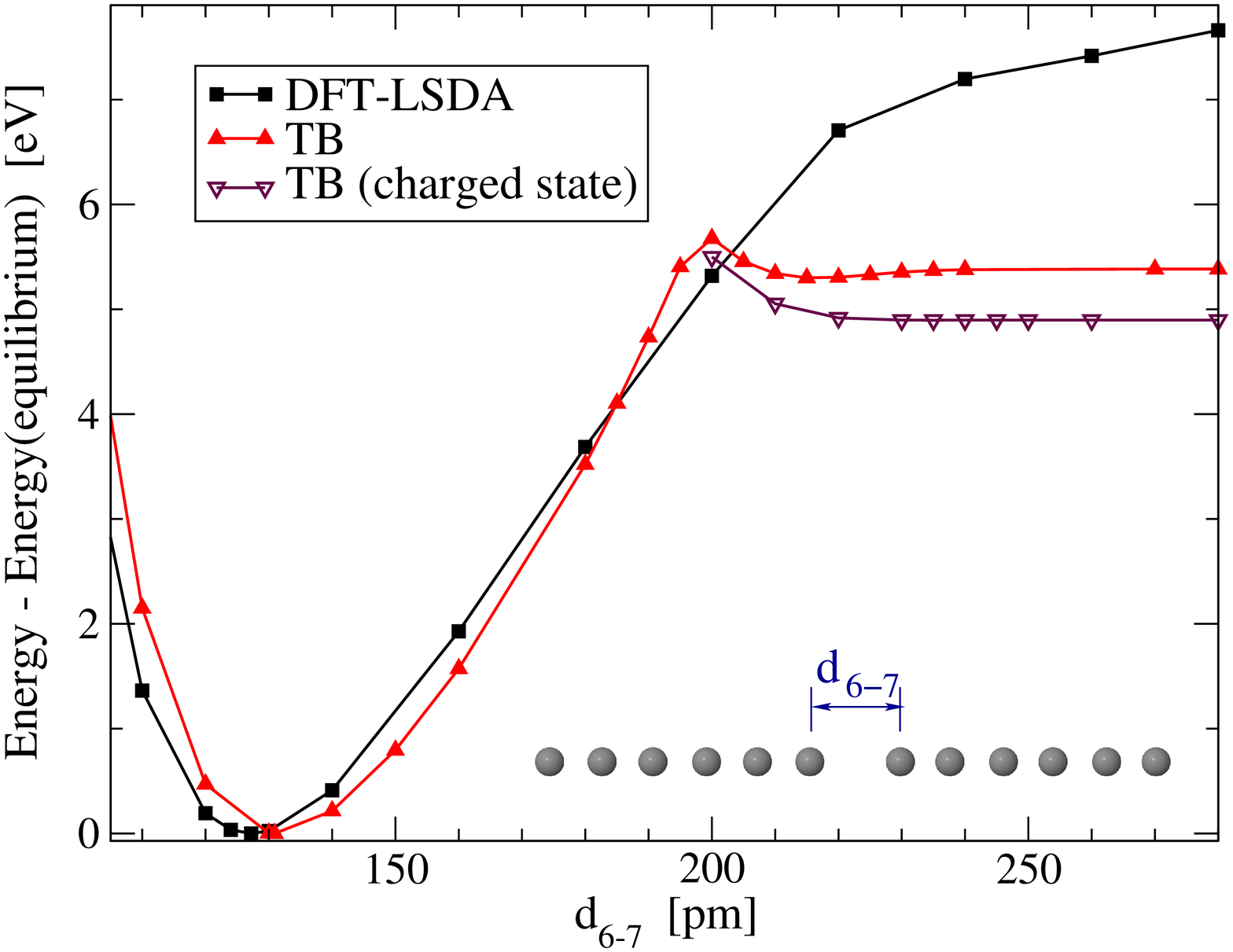} \hfill
\caption{\label{compare:fig}
A comparison of the adiabatic potential energy of a C$_{12}$ spCC as a
function of the length $d_{6-7}$ of the central bond, for a DFT-LSDA {\it
  ab-initio} model and the adopted TB model.
In both calculations, all other bonds in the spCC are fully relaxed at each
fixed elongation $d_{6-7}$ of the central bond.
The TB calculation of the charged state (down triangles) yields a
symmetry-broken configuration where two electrons localize on one C$_6$
unit, thus effectively realizing a ion pair C$_6^{2+}$--C$_6^{2-}$.
}
\end{figure}

The adopted parametric TB model \cite{Xu92} was constructed targeting
mainly $sp^3$ and $sp^2$ carbon.
Its adoption to describe spCCs should therefore be validated.
To this purpose, we perform {\it ab-initio} DFT-LSDA simulations of a
stretched carbyne.
We adopt a plane-waves basis (cutoff: 30 Rydberg) and ultrasoft
pseudopotentials to account for core electrons, as implemented in the
Quantum Espresso code \cite{espresso2009}.
To test the reliability of the adiabatic potential resulting in the TB
model, we compare the the TB total-energy adiabatic surface with the one
obtained by DFT-LSDA for a C$_{12}$ chain.
In particular, we consider the variation of the adiabatic energy as the
central bond between atoms 6 and 7 is stretched, with all other bonds fully
relaxed.
Figure~\ref{compare:fig} reports this comparison.
The minimum energy is realized at a very similar bond length, within 4~pm,
and the curvature in the minimum region is quite comparable.
The main deviations occur when the central bond is stretched beyond 200~pm,
approaching the cutoff distance (260~pm) of the TB model.
The dangling bonds at the C$_{12}$ ends produce a magnetic (spin-1) DFT
ground state.
The corresponding TB electronic structure exhibits a twofold-degenerate
molecular orbital (half-)occupied by 2 electrons at the Fermi level.
As the 6-7 bond is stretched, another level moves closer, until a further
degeneracy is realized accounting for the two half-filled $\pi$ orbitals of
the two identical C$_6$ units.
For $d_{6-7}\geq 200$~pm, this degeneracy can be lifted by charge transfer
of two electrons from one C$_6$ fragment to the other, with a corresponding
symmetry breaking and different bonding pattern of the two ions.
This distortion leads to energy lowering, as illustrated in
Fig.~\ref{compare:fig}. 
This distortion however is an artifact of the TB model, which allows it
because the TB model lacks any realistic description of the electrostatic
charging energy.
No sign of this artifact is observed in the DFT model, which instead
switches from a total-spin $S=1$ to a $S=2$ state as $d_{6-7}$ exceeds
$\simeq 220$~pm.

Despite the risk of charging artifacts such as the one described above, the
lack of magnetic exchange, and the short range, the overall shape of the
adiabatic potential is essentially satisfactory to the purpose of the
present calculations.
The bond-breakage energy is approximately 5.5~eV, not too far from the DFT
predicted value of approximately 8~eV.
This underestimation of the binding energy, in turn, suggests that the
stability of carbynes and thus the tendency to form them in real life
exceeds the one predicted by the present model simulations.

\section*{References}


\end{document}